\documentclass[acmtog,authorversion,nonacm]{acmart}

\usepackage[most]{tcolorbox}
\lstdefinestyle{jsonstyle}{
    basicstyle=\footnotesize\ttfamily,
    breaklines=true,
    frame=single,
    backgroundcolor=\color{gray!10},
    keywordstyle=\color{blue},
    stringstyle=\color{red},
    commentstyle=\color{green!70!black},
}
\usepackage{comment}
\AtBeginDocument{%
  }

\setcopyright{acmlicensed}
\copyrightyear{2025}
\acmYear{2025}
\acmDOI{XXXXXXX.XXXXXXX}
\acmConference[Conference acronym 'XX]{Make sure to enter the correct
  conference title from your rights confirmation email}{June 03--05,
  2018}{Woodstock, NY}
\acmISBN{978-1-4503-XXXX-X/2018/06}

\acmSubmissionID{1547}

\begin{document}

%\title{ReelWave: A Multi-Agent Framework Toward Professional Movie Sound Generation}

\title{ReelWave: Multi-Agentic Movie Sound Generation through Multimodal LLM Conversation}

\author{Zixuan Wang}
\email{zwanggk@connect.ust.hk}
\affiliation{%
  \institution{The Hong Kong University of Science and Technology}
  \country{Hong Kong}
}

\author{Chi-Keung Tang}
\email{cktang@cse.ust.hk}
\affiliation{%
  \institution{The Hong Kong University of Science and Technology}
  \country{Hong Kong}
}

\author{Yu-Wing Tai}
\email{yu-wing.tai@dartmouth.edu}
\affiliation{%
  \institution{Dartmouth College}
  \country{USA}
}

% \affiliation{%

% \affiliation{%

% \affiliation{%

% \affiliation{%

% \affiliation{%

% \affiliation{%

% \affiliation{%


\begin{abstract}
Current audio generation conditioned by text or video focuses on aligning audio with text/video modalities.  Despite excellent alignment results, these multimodal frameworks still cannot be directly applied to compelling movie storytelling involving multiple scenes, where `on-screen' sounds require temporally-aligned audio generation, while `off-screen' sounds contribute to appropriate environment sounds accompanied by background music when applicable. Inspired by professional movie
production, this paper proposes a multi-agentic framework for audio generation supervised by an autonomous Sound Director agent, engaging multi-turn conversations with other agents for on-screen and off-screen sound generation through multimodal LLM. 
%, see Figure~\ref{fig:teaser}. 
To address on-screen sound generation, after detecting any talking humans in videos,
%we first examine if any person is talking on the screen. If so, we use a pretrained lip-to-speech model to generate on-screen voice. Otherwise, 
we capture semantically and temporally synchronized sound by training a prediction model that forecasts interpretable, time-varying audio control signals: loudness, pitch, and timbre, which are used by a  Foley Artist agent to condition a cross-attention module in the sound generation. The Foley Artist works cooperatively with the Composer and Voice Actor agents, and together they autonomously generate off-screen sound to complement the overall production. Each agent takes on specific roles similar to those of a movie production team. To temporally ground audio language models, in ReelWave, text/video conditions are decomposed into atomic, specific sound generation instructions synchronized with visuals when applicable. Consequently, our framework can generate rich and relevant
audio content conditioned on video clips extracted from movies.             
\end{abstract}





\begin{teaserfigure}
\begin{center} \includegraphics[width=0.80\linewidth]{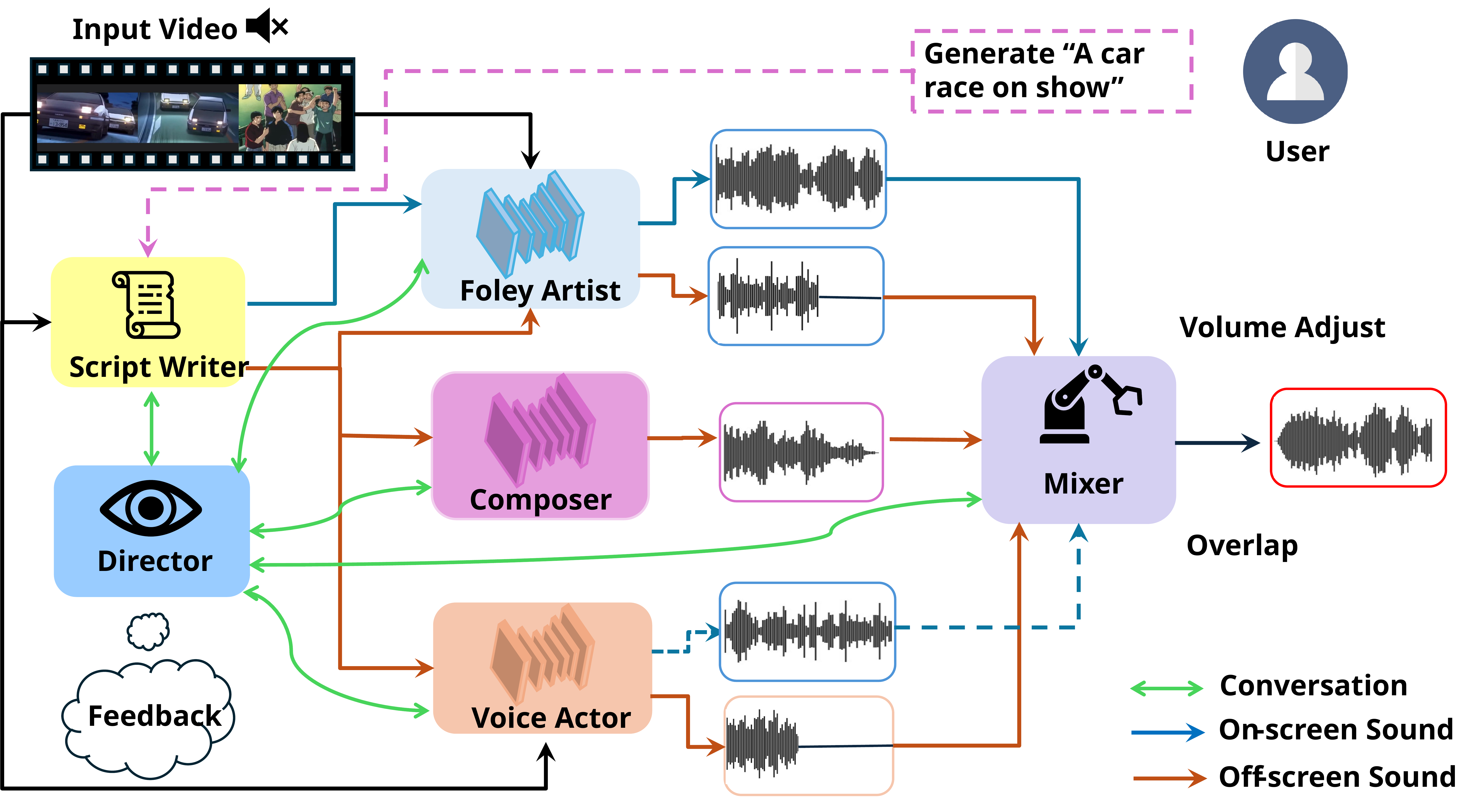}
\end{center}
\caption{ReelWave is a multi-agentic framework for automatically generating visually aligned audio given video clips, coupled with an optional module when the user input is a plot summary, indicated by the dashed line.}
\label{fig:teaser}
\end{teaserfigure}

\maketitle

%\begin{comment}
\begin{figure*}[tp]
\begin{center}
\includegraphics[width=0.90\linewidth]{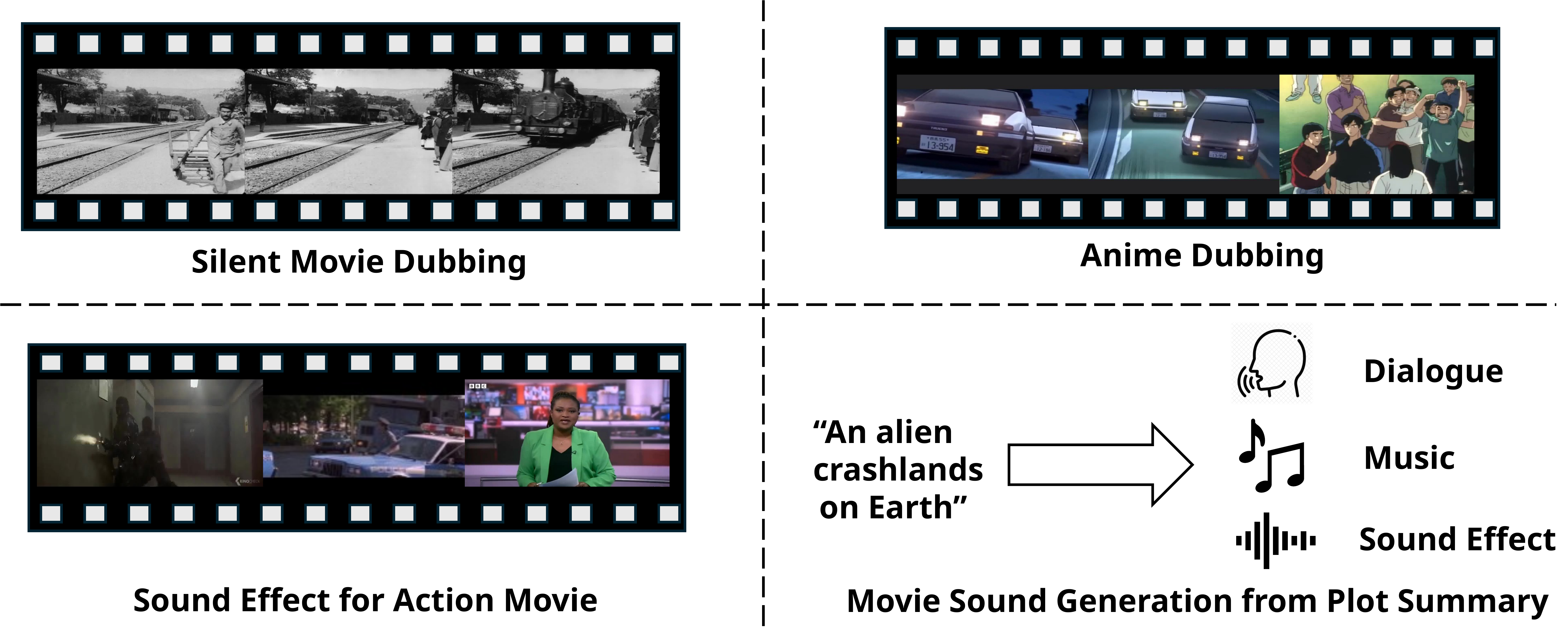}
\end{center}
    \caption%of{figure}
{\textbf{ReelWave} is applied to generate sound for different scenarios. Our multi-agent framework takes a short plot summary description and generates sound effects, audible dialogues, and music that are suitable for the plot. Please review the supplemental videos with audio.}    \label{fig:figure2}
\end{figure*}
%\end{comment}

\section{Introduction}
\label{sec:intro}

% To insert a figure: \input{figs/template}
% Or table: \input{tables/template}

Following text-guided audio generation, more research attention has been paid to video-conditioned audio generation, specifically, on aligning video and audio modalities. 
Despite excellent results, 
state-of-the-art synchronized audio still falls short of meeting the requirement for real-world sound production for video clips much less for movies and films, for instance. %, see Figure~\ref{fig:figure2}. 
For more immersive visual storytelling, sound design goes beyond synchronized sound effects to include music and dialogue. `On-screen' sound effects, either human speech or Foley sounds, are those that need precise synchronization with the visual action. On the other hand, `off-screen' sound contributes to the overall context of the scene, such as dialogue, environment sounds, or background music. Additionally, generating visual sound without recognizing multiple scenes often leads to suboptimal performance, as noted in~\cite{yi2024efficient}, failing to meet the increasing demand for generative audio in longer, multi-scene videos.

Recent advances in large language models have made it possible to use LLM for task planning and carrying out a variety of downstream tasks, including the planning and execution of robot tasks and general human motion generation. %~\cite{wu2024motion}. 
Confronted with the complexities of real-world tasks, works such as CAMEL~\cite{li2023camel} proposed a communicative agent framework that performs task planning with multiple rounds of conversation. Although audio generation can potentially involve complexity in planning and generation, particularly when the condition incorporates multiple events in text or multiple scenes in video, most of the current work only focuses on single-round inference, with few involving LLM in explicit planning and none in providing feedback and revision in audio generation.

In this paper, we introduce ReelWave, a multi-agent framework for movie sound generation. As shown in Figure~\ref{fig:teaser},  our framework draws inspiration from the real filmmaking process. We first use a training-free scene detector to divide the conditional video into multiple scenes, then utilize a video understanding model to obtain the textual description of the visual scene. To produce aligned on-screen sound, we either utilize a lip-to-speech model to generate speech or a prediction model to predict three interpretable time-varying audio control signals of the to-be-generated sound from video, namely loudness, pitch, and timbre. To complement the generation with off-screen sound, we utilize a role-playing setting that incorporates a sound director who supervises the generation process, a voice actor for dialogues, a composer to compose music, and finally, a mixer to mix multiple generated soundtracks with the correct sequence and reasonable volume. Our contributions are summarized as:

\begin{itemize}
    \item We propose ReelWave, a multi-agent {\em role-playing} framework for movie sound generation inspired by filmmaking production. We further define two methods for on-screen sound and off-screen sound generation.
    \item ReelWave can generate {\em synchronized} on-screen audio from video input featuring several scenes based on three explicable audio qualities derived from the video.
    \item ReelWave can further complement the generation with {\em dialogues and music} through a multi-agent framework with role-playing and automatic feedback.
\end{itemize}

To the best of our knowledge, our framework is among the first to introduce multiple, autonomous sound making agents operating in a multimodal conversational LLM framework for multi-scene audio generation. We believe ReelWave will contribute to sound design artists and editors by providing a more expressive and interpretable generation tool. We highly encourage the reader to view and listen to our audio examples in the supplemental video, Figure~\ref{fig:figure2}.

\section{Related Work}
\label{sec:related}

\noindent \textbf{LLM-based Agent Method.~}
Recent advancements in large language models (LLMs) have fostered significant progress in using LLMs for task planning and execution across various domains, such as robot task planning and software development~\cite{driess2023palm, rawles2024androidinthewild, yang2023appagent}. LLMs demonstrate the ability for zero-shot or few-shot generalization, enabling efficient transfer of tasks without the need for significant changes in the model parameters~\cite{xi2023rise}. In the domain of audio generation, WavJourney~\cite{liu2024wavjourney} and WavCraft~\cite{liang2024wavcraft} pioneered the use of LLMs to connect different audio models for creative audio production. However, these systems are limited to a single-agent setting, which may fall short when the task becomes more complex and do not leverage the full potential of GPT-4’s capabilities, often resulting in audio that doesn't meet user-specified length or complexity requirements.

To enhance system capacity, %complexity into the system, 
CAMEL~\cite{li2023camel} explores the use of a multi-agent framework to divide and conquer tasks related to math, coding, and science, with each agent adopting the role of either an AI Assistant or AI User. This method improves scalability and ensures the conversation follows the right direction. Inspired by this work, we introduce a supervisor agent over multiple planning and generation agents. The supervisor agent coordinates the generation of the audio components (music, dialogue, soundtrack) in a collaborative manner, providing feedback across multiple rounds if the initial plans prove unsatisfactory. This mirrors the real-world process of movie sound production, where the sound director continuously works with the sound team to adjust the generated audio based on evolving creative needs.

\vspace{1mm}

\noindent \textbf{Diffusion-based Audio Generation.~}
The introduction of the latent diffusion method for audio generation was a significant breakthrough in the field, with works like AudioLDM~\cite{liu2023audioldm} leading the way. Subsequent models, such as Tango~\cite{ghosal2023text}, AudioLDM-2~\cite{liu2024audioldm}, and Auffusion~\cite{xue2024auffusion}, have used pre-trained LLMs such as Flan-T5~\cite{flanT5} for efficient text encoding, demonstrating high-quality results through large-scale training. These text-conditioned methods have opened the door to using visual input for audio generation, offering a more efficient alternative to training video-to-audio generators from scratch by augmenting existing audio generation frameworks with visual conditioning~\cite{wang2024audio}.

FoleyCrafter~\cite{zhang2024foleycrafter} pushes the boundaries by incorporating additional modules to adapt visual semantics, detect timestamps, and align temporal conditions for generating audio. ReWaS~\cite{jeong2024read} further advances this by integrating continuous timestamp conditions, though it still lacks consideration of visual semantic style and relies mainly on input captions. Our multimodal LLM conversational framework builds upon these technical contributions, by introducing a comprehensive multi-agent setup to model both the ``on-screen'' and ``off-screen'' audio conditions, enabling a more nuanced and coherent audio generation process for longer video segments. In particular, our approach allows for the exploration of multiple scenes, which is a common characteristic of real-world movie productions.

\vspace{2mm}

\noindent \textbf{Integration of Visual and Temporal Conditions for Audio Generation.~}
Text-to-audio (TTA) generation has seen significant progress with models such as Make-an-Audio 2~\cite{huang2023make}, which was among the first to handle complex textual descriptions for audio generation. However, this work primarily focuses on  sequencing of generated audio and lacks the flexibility for multi-turn editing. Additionally, Make-an-Audio 2 is limited by extensive training with LLM-based data augmentation which restricts its applicability to TTA tasks. 

For video-to-audio (V2A) generation, works such as FoleyGen~\cite{mei2024foleygen} and V2meow~\cite{su2024v2meow} have employed transformers to model the audio generation process. However, these autoregressive models tend to suffer from accumulating errors during inference, which can degrade the quality of long-form audio generation~\cite{li2024autoregressive}. Our method addresses this limitation by using a multi-agent framework, which provides iterative feedback and enhances the stability of long-form, multi-scene audio generation, ensuring higher consistency and quality over time.

Furthermore, our framework incorporates both visual and temporal conditioning in a unified model, leveraging the rich context provided by multiple scenes to generate more sophisticated and contextually accurate audio. This differs from previous methods that rely primarily on textual descriptions and struggle with generating audio of the desired length or quality. By handling multi-scene video clips, our model can generate audio that is semantically coherent and aligns well with the temporal structure of movie production, thereby facilitating more realistic and professional-level sound design.

\vspace{2mm}

\noindent \textbf{Comparison to Previous Work.~}
Our approach offers several advantages over previous methods. Firstly, the incorporation of a multi-agent framework with a dedicated Sound Director allows for iterative refinement and collaboration throughout the audio generation process. This is a key feature that distinguishes our work from methods that rely on a single-stage generation process or those lacking feedback mechanisms. Secondly, our framework is designed for long-form, multi-scene audio generation, enabling it to handle more complex and context-rich audio tasks that are typical in movie production. Lastly, by leveraging both visual and temporal conditions in a unified model, we ensure that our generated audio is not only contextually relevant but also richly detailed, matching the demands of professional movie sound design. 

The framework closest to ours is ComposerX~\cite{deng2024composerx}, which proposes using a multi-agent framework to generate symbolic music notation for melody, 
harmony, and instrumentation. On the other hand, pure music generation is a far more structured and limited domain compared to general video/text-conditioned audio generation, where the latter demands temporal precision and semantic diversity. The complexities involved in aligning audio to the real-world video, understanding cross-modal relationships, and modeling a multi-agent system to generate audio pose extra challenges beyond those for generating music components.

\section{Method}
\label{sec:method}

ReelWave is a multi-agent framework that consists of a two-way agent communication network between the a)~supervisor agent, namely as Sound Director, and b)~planning and generation agents, namely as Script Writer, Foley Artist, Composer, Voice Actor, and Mixer. 

The on-screen sound is produced either by the Foley Artist, which encompasses a sound prediction module given the video condition, or the Voice Actor, which consists of a pretrained lip-to-speech module. To further enrich the generation content, additional off-screen sound is produced by the three agents presented in the middle of Figure~\ref{fig:teaser}. Considering single agent planning framework may have a limited perspective and thus a higher error rate, we utilize a supervisor agent to offer critique and guidance during the planning process. In addition to generating the accompanying audio from video clips, ReelWave may take a text-only plot summary from the user and generate multi-scene sound, denoted as the pink dash line in Figure~\ref{fig:teaser}. Through generating multiple overlapping soundtracks, we take a step further toward real-world movie sound generation. 

\subsection{Scene detection and understanding}
V2A-SceneDetector~\cite{yi2024efficient} proposed an efficient way to recognize scene boundaries within a given video by computing the similarity difference between consecutive frames. However, the authors did not fully demonstrate the potential of this method by limiting the experiments to the VGGSound dataset, where the majority are single-scene videos. Our framework employs this scene detector to partition a given movie-like video input into segments. Let 
%The method is explained as follows, where 
\(\mathbf{X} = [\mathbf{x}_1, \mathbf{x}_2, \ldots, \mathbf{x}_T] \in \mathbb{R}^{T \times D}\) represent the sequence of frame embeddings, where \(T\) is the number of frames, and \(D\) is the embedding dimension.
%\begin{enumerate}
%\item 
Then we compute pairwise similarities between frame embeddings followed by centering, where self-similarity matrix \(\mathbf{S} \in \mathbb{R}^{T \times T}\):
\begin{equation}
\mathbf{S}_{i,j} = \mathbf{x}_i \cdot \mathbf{x}_j \\ 
\end{equation}
%normalizing $\mathbf{S}$: by subtracting row and column means:
\begin{equation}
\mathbf{S}_{\text{centered}} = \mathbf{S} - \text{mean}(\mathbf{S}, \text{axis}=0) - \text{mean}(\mathbf{S}, \text{axis}=1)^{\top}
\end{equation}
%\item 
followed by computing absolute differences between consecutive rows:
\begin{equation}
    \Delta\mathbf{S}_i = \sum_{j=1}^T |\mathbf{S}_{\text{centered},i,j} - \mathbf{S}_{\text{centered},i+1,j}|
\end{equation}
and identifying significant peaks in $\Delta\mathbf{S}$ using a pre-defined threshold:
\begin{equation}
    \text{Peaks} = \{ i \mid \Delta\mathbf{S}_i > \tau \cdot \sigma_{\Delta\mathbf{S}} \}.
\end{equation}
%\item 
Finally, the scene boundary is defined as 
\begin{equation}
    \{(s_k, e_k)\} = \{(0, p_1 - 1), (p_1, p_2 - 1), \ldots, (p_n, T - 1)\}
\end{equation}
where $p_1,...,p_n$ are detected peak indices.
%\end{enumerate}

Afterward, we use a video understanding model~\cite{zhang2024videoinstructiontuningsynthetic} to capture the key event description and full description. The key event description reflects simple and atomic auditory elements encompassed in the video, such as the type of sound or the event causing it. For example, ``The sound of explosions'' and ``The sound of roaring engines.'' 
We use this concise description as the caption for on-screen sound production. The full description aims to capture the full ongoing event and visual environment, which provides the context for the framework to infer off-screen sound. 

In addition, the video understanding model checks whether there exists any talking human in the input video. If so, a pretrained lip-to-speech model is used for audible voice generation. We treat this as the on-screen sound. Otherwise, we proceed by passing the key event description to the Foley Artist for on-screen sound generation. Details are described in Section~\ref{agent-planning}.
\begin{comment}
    \textcolor{red}{If talking humans in the video are detected, a pretrained lip-reading model is used to infer the corresponding transcription, which will be forwarded as an audio script to the pertinent agent for audible voice generation.} 
\end{comment}

\subsection{Multi-agentic generation and planning}\label{agent-planning}
\noindent\textbf{Script Writer.~}
The full description of the video is passed to the Script Writer to convert into a structured script, similar to the storyboard used in the movie industry. The Script Writer aims to infer the storyline behind the visual scene, in addition to the environmental setting suitable for the story to happen. The script consists of five key elements: 
\vspace{2mm}
\begin{itemize}

  \item \textit{Location Description}: description of the event location
  \item \textit{Direction Notes}: vocal style and tone for the main character and the surrounding environment. 
  \item \textit{Plot Summary}: an overview of the scene content
  \item \textit{Character}: characters involved in the scene.
  \item \textit{Dialogue Script}: dialogue lines for each character involved. To provide context as much as possible, the agent may also specify voice tone in round brackets and action/setting description in square brackets.

\end{itemize} 
%\vspace{2mm}
 Figure~\ref{fig:script} shows an example script. The script is then passed as the main reference for the subsequent plan generation.
 
\begin{figure}[t]
\begin{tcolorbox}[title=Script, fonttitle=\bfseries\small, coltitle=black, colbacktitle=gray!20, colback=white, colframe=gray!50, boxsep=1mm, left=1mm, right=1mm, top=1mm, bottom=1mm, fontupper=\footnotesize]
    SCENE 1 \\
    
    Location Description: Inside of a spaceship. \\
    Direction Notes: A male vocal style for THE ALIEN. A female voice for COMPUTER SYSTEM \\
    Plot Summary: A young alien crashlands on Earth. \\
    Character: THE ALIEN; COMPUTER SYSTEM \\
    
    Dialogue Script: \\
    THE ALIEN \\
    \lbrack In the cockpit of the spaceship] \\
    ``Whoa! Can't wait to get off this planet. Ain't nothing that can stop me now! " \\
    
    COMPUTER SYSTEM \\
    ``Warning, navigation systems are offline." \\
    
    THE ALIEN \\
    \lbrack Speaking to the on-board AI\rbrack \\
    ``No! No, no, no, no, no, this can't be happening! \\
    \lbrack Spaceship hurtles through the atmosphere before crashing into a scrap yard\rbrack \\
    
    END SCENE 1
\end{tcolorbox}
\caption{Script example.}
\label{fig:script}
\end{figure}

\vspace{2mm}
\noindent\textbf{Foley Artist.~}
Foley Artist consists of two parts: on-screen Foley sound generator and off-screen Foley sound planner. We define on-screen Foley sound as the main events that happen on the screen, which is generated using the key event description from the video understanding model, with {\em Loudness, Pitch, Timbre}, as the conditions for producing synchronized sound. Further details are illustrated in Section~\ref{subsec:conditioning}.

Off-screen Foley sound corresponds to environmental sound that illuminates the event's ambiance, such as ``distant screams of townspeople'' or ``occasional distant gunfire'' in a war scene. The script is provided to the off-screen Foley sound planner. To better complement the on-screen Foley sound, we provide the key event description together with the script. The agent is asked to focus on generating ambient or event sounds while excluding human voices and music. In addition, the description for each soundtrack should be simple and atomic, directly describing the type of sound or the event making the sound. The output of the sound planner is a detailed Foley design plan for off-screen sound generation. 
\textcolor{black}{See Figure~\ref{fig:foley_plan} 
for sample JSON Foley plan generated.}

\vspace{2mm}
\noindent\textbf{Composer.~}
The task for the Composer is to translate the audio script into a detailed music design plan for generation 
\textcolor{black}{(see 
Figure~\ref{fig:music_plan} in 
supplementary material 
for music plan).} % . 
The agent is asked to focus on generating {\em only} the music, with the description capturing the style and instruments for each soundtrack. If the agent thinks music is unnecessary, it will respond N/A. Only the script is provided for reference.

\vspace{2mm}
\noindent\textbf{Voice Actor.~}
For off-screen voice generation, the Voice Actor will translate the audio script 
% \textcolor{red}{(which may be user input or directly deduced by lip reading from the silent video)} 
into a detailed voice design plan for generation. 
\textcolor{black}{See Figure~\ref{fig:voice_plan} in supplementary material 
for voice plan.} The agent focuses {\em only} on generating the human voice, with one dialogue for each soundtrack. To stylize %further add variety for 
the generation, either a male voice or a female voice can be specified. We additionally add a total word constraint for every dialogue, assuming two words per second, which is close to human speaking speed in normal conversation. Similar to the Composer agent, only the script is provided, with the agent responding N/A if the scene does not have human dialogues.

For on-screen voice generation, in the case when the video understanding model detects a person talking, the agent will use a pretrained lip-to-speech model to directly generate human voice from input silent video with lip synchronization.

\vspace{2mm}
\noindent\textbf{Mixer.~}
After obtaining the textual plan from the Foley Artist, Composer, and Voice Actor, the Mixer will conduct a final mix of multiple soundtracks to ensure the right balancing and time order
\textcolor{black}{
(see  Figure~\ref{fig:mixer_plan} %supplementary material 
for mixer plan)}. 
Specifically, it will assign appropriate volume and adjust the start time and end time to each soundtrack. The volume is an integer value in dB following the LUFS standard. For background sound effect/music, the usual level is around $-35$ to $-40$ dB. For speech is around $-15$ dB.

\vspace{2mm}
\noindent\textbf{Sound Director.~}
The sound director conducts a two-way conversation with the Foley Artist, Composer and Voice Actor, providing feedback and instructions to refine the generation plan, Figure~\ref{fig:conversation}. In case any inconsistency happens between the agent's action and the task, the director will provide feedback to rectify the agent's behavior. This is similar to the movie industry, where the soundtracks are not finalized in one go but rather through multiple rounds of refinement between the production team and the director. In our framework, we conduct multiple rounds of conversation between individual generation agents and the director. The conversation ceased when the director output $\langle\textit{TASK\_DONE}\rangle$ token. We prompt the director to enrich the design plan with more details and instruct the planning agent to add additional soundtracks when appropriate. In the case when the input is a plot summary, the director may ask the Script Writer to add additional scenes if necessary. 

% \textcolor{red}{is it possible that individual planning agents converse among each other without the director??}

\begin{figure}[t]
\begin{tcolorbox}[ fonttitle=\bfseries\small, coltitle=black, colbacktitle=gray!20, colframe=gray!50, boxsep=1mm, left=1mm, right=1mm, top=1mm, bottom=1mm, fontupper=\footnotesize]
\textcolor{red}{User}: **Input a video of war scene** \\

%\textcolor{cyan}{Script Writer}: \\
%\textit{(JSON plan describing the script)} \\
\textcolor{blue}{Sound Director}: \\
Instruction: Please add a specific sound effect description for the projectiles before they hit, such as a whistling or swooshing sound, to enhance the anticipation and impact of the explosions. \\
\textcolor{cyan}{Script Writer}: \\
\textit{(Revised script)} \\
\textcolor{blue}{Sound Director}: TASK\_DONE.\\

%\textcolor{cyan}{Foley Artist}: \\
%\textit{(JSON plan for Foley sound generation)} \\
\textcolor{blue}{Sound Director}: \\
Instruction: Add a soundtrack of the sound of fiery explosions illuminating the scene, with bursts of orange and yellow flames, to enhance the visual description of the siege. Set this sound to start at 1 second and end at 3 seconds. \\
\textcolor{cyan}{Foley Artist}: \\
\textit{(Revised Foley sound plan)} \\
\textcolor{blue}{Sound Director}: TASK\_DONE.\\

%\textcolor{cyan}{Composer}: \\
%\textit{(JSON plan for music generation)} \\
\textcolor{blue}{Sound Director}: \\
Instruction: Add a second soundtrack starting at 3 seconds and ending at 6 seconds with a style of "Choral, Ominous" and a description that captures the atmosphere of the castle under siege, focusing on the emotional impact of the battle on the surroundings.. \\
\textcolor{cyan}{Composer}: \\
\textit{(Revised music plan)} \\
\textcolor{blue}{Sound Director}: TASK\_DONE.\\

\end{tcolorbox}
\caption{\textbf{Conversation among Sound Director and planning agents.} This example shows the Director provides feedback and suggestions according to the generated plan, which is refined through multi-round conversations. The conversation between the Sound Director and Voice actor or Mixer follows a similar structure, which we omitted due to space limitations.}
\label{fig:conversation}
\end{figure}

\vspace{2mm}
\noindent\textbf{Output structure.}
All communication is operated through text, and the agents will output a plan in JSON format; see supplementary material. The JSON file is further used for the generation of backbones to generate audio soundtracks. Our design offers a transparent, structured way for movie sound generation that is easy to monitor. In addition, the user may optionally provide the custom dialogue lines to suit his or her needs. Please see the supplementary for JSON examples for Foley plan, music plan, and voice plan.

\begin{figure*}[tp]
    \centering
    \includegraphics[width=0.80\linewidth]{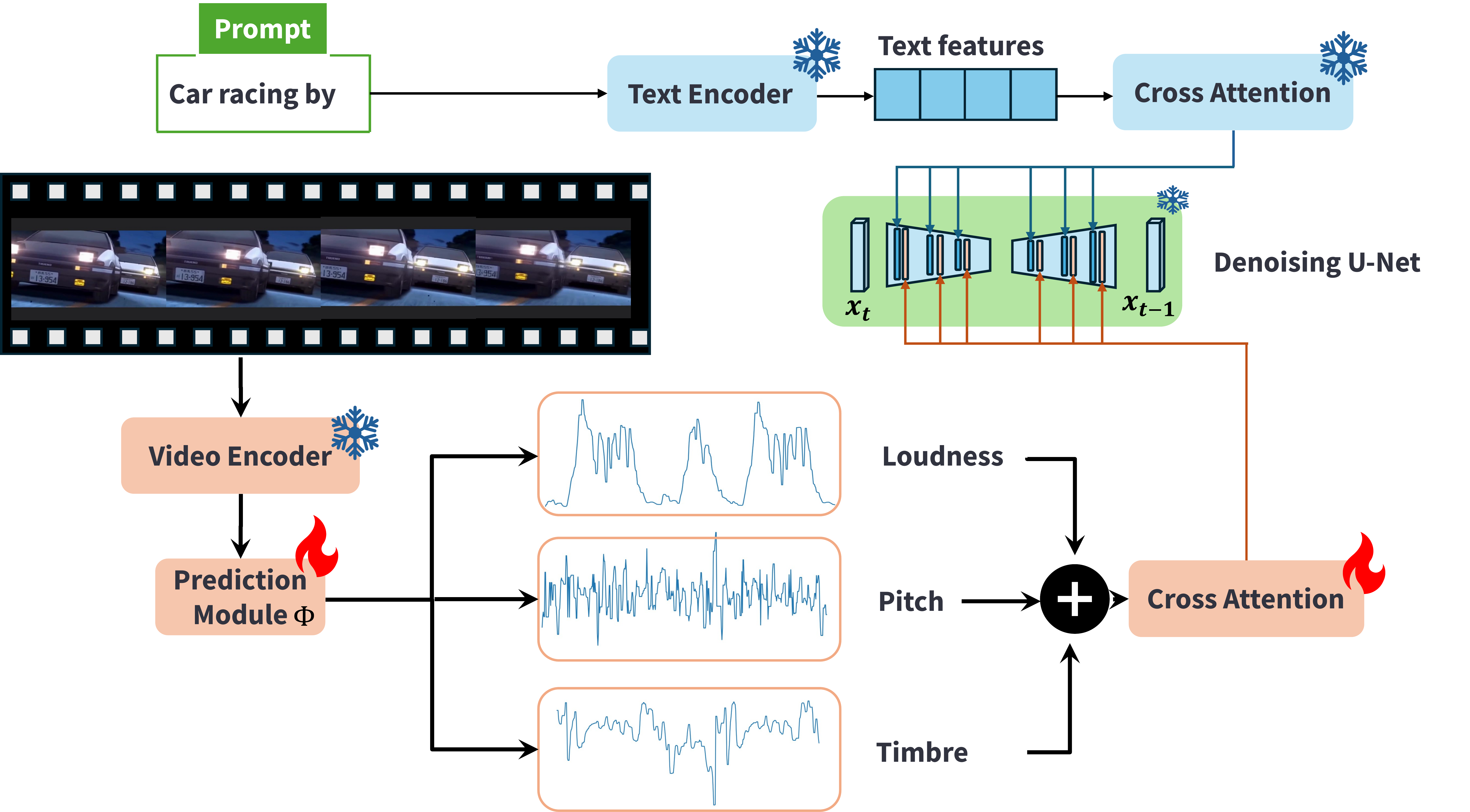}
    \caption{\textbf{ReelWave’s conditioning method for on-screen sound generation}. ReelWave uses a prediction model to extract loudness, pitch, and spectral centroid (timbre), which further serve as conditions for generation via cross-attention layers inside the diffusion UNet.}
    \label{fig:figure3}
\end{figure*}

\subsection{Audio control signals prediction from video} 
\label{subsec:conditioning}
We use three time-varying interpretable signals, namely Loudness, Pitch, and Timbre, as the control signals for synchronized on-screen sound generation, drawing inspiration from Sketch2Sound~\cite{garcia2024sketch2sound}, where these sound properties have been shown to achieve impressive results on audio style control and transfer. We further extended the application to video-conditioned audio generation, made adjustments in our implementation compared to the original paper, as 
Sketch2Sound is not open-sourced at the time of writing. 
The following summarizes how we extract these properties:

\vspace{2mm}
\begin{itemize}

  \item \textit{Loudness}: We extract the magnitude spectrogram using Short-Time Fourier Transform (STFT) and compute the root mean square loudness of the A-weighted signal across time frames.

  \item  \textit{Pitch}: We extract the raw pitch probabilities of using CREPE (``tiny” variant from PyTorch implementation\footnote{https://github.com/maxrmorrison/torchcrepe})~\cite{kim2018crepe} pitch estimation model. We further zero out all probabilities below 0.1 in the pitch probability matrix, and finally decode back to the pitch signal. 

  \item  \textit{Timbre}: We use the center of mass of the frequency spectrum per audio frame (Spectral Centroid) as the measurement of timbre. Audio that has a brighter timbre will contain frames of a higher spectral centroid. Following~\cite{garcia2024sketch2sound} we extract and convert the spectral centroid to a MIDI-like representation.
\end{itemize}
\vspace{2mm}

Figure~\ref{fig:figure3} illustrates the conditioning mechanism:
After extracting the audio properties, we concatenate them, resulting in a conditional signal of shape $a \in \mathbb{R}^{L \times 3}$, where $L$ denotes the sequence length.

We train a prediction model to predict the control signals from the video features. The prediction model contains two transformer blocks, three linear blocks, and three prediction heads, where the prediction heads share the same input features. Considering the different lengths between the video features and the control signals, we use interpolation to reshape the ground truth signals. During inference, we again use interpolation to obtain the predicted signals. 

We 
%follow ~\cite{jeong2024read} to 
use MSE loss when training the model. To stabilize the fluctuation of the conditional signal between each temporal window, we apply a median filter on the ground truth signal during training. During inference, no median filter is used, and we empirically find that it increases the performance. 
As noted by ~\cite{jeong2024read}, conventional approaches that use CLIP~\cite{radford2021learning} for video encoding may result in suboptimal performance on temporal alignment. We employ SynchFormer~\cite{iashin2024synchformer} to extract the video features, which is designed and trained to encode motion dynamics and semantics. We freeze the video encoder during training.
The learning objective is:
\begin{equation}
\mathcal{L}_{ctrl} =||\phi(E_v) - \texttt{Resize}(\texttt{Midfilt}({a}))||_2^2
\end{equation}
where $\phi$ denotes the prediction model that we train, $E_v$ denotes the extracted video feature from SynchFormer, and $a$ denotes the conditional signal.

\subsection{Signal-conditioning for on-screen sound} 
The generation backbone for on-screen sound consists of a pretrained text-to-audio diffusion model. We further apply cross-attention to apply the control signal to the generation process. During the training, we optimize the visual-based cross-attention layers while freezing the original text-based cross-attention layers. 

The cross-attention mechanism is detailed as follows: Given a query feature $\mathbf{Z}$, text features $c_{txt}$, and control signals $c_{ctrl}$, the output for combining two types of cross-attention is defined as:

\begin{equation}
\begin{split}
\!\!\!\!\!\!\mathbf{Z}^{new}=\text{Softmax}(\frac{\mathbf{Q}\mathbf{K}_{txt}^{\top}}{\sqrt{d}})\mathbf{V}_{txt}+\text{Softmax}(\frac{\mathbf{Q}(\mathbf{K}_{ctrl})^{\top}}{\sqrt{d}})\mathbf{V}_{ctrl}\\
 \noindent\text{where}\ \mathbf{Q}=\mathbf{Z}\mathbf{W}_{txt}^q,\mathbf{K}_{txt}=\boldsymbol{c}_{txt}\mathbf{W}_{txt}^k, \mathbf{V}_{txt}=\boldsymbol{c}_{txt}\mathbf{W}_{txt}^v, \\
\mathbf{K}_{ctrl}=\boldsymbol{c}_{ctrl}\mathbf{W}_{ctrl}^k, \mathbf{V}_{ctrl}=\boldsymbol{c}_{ctrl}\mathbf{W}_{ctrl}^v
\end{split}
\end{equation}

The text feature is obtained from the key event description using the original text encoder. In short, we only add two parameters $\mathbf{W}_{ctrl}^k$ and $\mathbf{W}_{ctrl}^v$ on top of each text cross-attention layer, and we use the same query for control signal cross-attention as for text cross-attention. During inference, we introduce another parameter $\lambda$ for balancing the guidance from text and control signal:
\begin{eqnarray}
\!\!\!\!\mathbf{Z}^{new}& =& \text{Attention}(\mathbf{Q},\mathbf{K}_{txt},\mathbf{V}_{txt}) + \nonumber \\
& & \lambda\cdot\text{Attention}(\mathbf{Q},\mathbf{K}_{ctrl},\mathbf{V}_{ctrl}).
\end{eqnarray}

The final objective for the diffusion process, which is similar to latent diffusion models, is
\begin{equation}
L_{\text{diffusion}}=\mathbb{E}_{\boldsymbol{x}_{0},\boldsymbol{\epsilon}, \boldsymbol{c}_{txt}, \boldsymbol{c}_{ctrl}, t} \| \boldsymbol{\epsilon}- \boldsymbol{\epsilon}_\theta\big(\boldsymbol{x}_t, \boldsymbol{c}_{txt}, \boldsymbol{c}_{ctrl}, t\big)\|^2.
\end{equation}

\subsection{Implementation Details}
During training, we randomly drop the text feature and control signal with equal probability of 0.05, following~\cite{ye2023ip}. We train separately the prediction model and the conditional layers. During inference, we set $\lambda=0.5$ as default.

Since the video encoder that we use, SynchFormer, is trained on 4.8-sec clips, our training and evaluation are conducted on 5-sec videos. In our multi-agentic, multimodal LLM conversation framewkork, the generation backbone for Foley Artist is Auffusion~\cite{xue2024auffusion} for on-screen sound, and Stable-Audio-1.0~\cite{evans2024stable} for off-screen sound. On the other hand, the backbone for Voice Actor is  LipVoicer~\cite{yemini2023lipvoicer} (for on-screen sound) and Bark~\cite{bark} (for off-screen sound). The backbone for Composer is MusicGen~\cite{copet2023simple}. For the video understanding model, we employ LLaVA-Video-7B-Qwen2~\cite{zhang2024videoinstructiontuningsynthetic} for its comparably compact size and powerful performance. We use GPT-4 Turbo~\cite{achiam2023gpt} in our multi-agentic system.

\section{Experiments}
\label{sec:experiments}

\subsection{Datasets}
We use AVSync15~\cite{zhang2024audio} for training and evaluation for the on-screen Foley sound generation module. AVSync15 is a curated dataset from VGGSound Sync~\cite{chen2021audio} that has 1500 high video-audio alignment pairs, which is ideal for training and demonstrating temporal alignment between video and audio. We repeat and concatenate all the videos into 5-sec videos and extract the audio ground truth at 16000 Hz.

\subsection{Evaluation Metrics}
We employ a comprehensive set of evaluation metrics to evaluate the audio quality and the synchronization for on-screen Foley audio generation. We use the Frechet distance (FD) and Mean KL Divergence (MKL) to evaluate audio fidelity. FD evaluates distributional similarity to evaluate the fidelity of the synthesized audio. MKL measures the paired sample-level
similarity using mean KL-divergence. To quantify audio-video relevance, we utilize CLIP Score and AV-Align. The CLIP Score is obtained using Wav2CLIP as the audio encoder and CLIP as the video encoder, which quantifies the similarity between the input video and the synthesized audio embeddings in the same representation space. 

To evaluate synchronization, we employ \# Onset Accuracy, Onset AP, Energy MAE, and AV-Align. AV-Align is based on detecting energy peaks in audio-visual modalities, The onset signal is obtained by thresholding the amplitude gradient. As noted in~\cite{jeong2024read}, these metrics may ignore relatively quiet sound effects or natural sounds. We further follow~\cite{jeong2024read} to report the mean absolute error (MAE) of the energy signals and AV-align score, which is based on detecting energy peaks in audio-video pairs. 

\subsection{Evaluation and comparison}
Table~\ref{tab:semantic} shows that for semantic alignment and audio quality, our model outperforms other baselines on MKL, while consistently ranking second on other metrics. This is a strong indication that ReelWave can successfully predict audio attributes inherent in the given video, offering an explainable temporal and semantic meaning. With only training the model on a small dataset (1,500 samples), our model achieves a reasonable result, while in comparison, Seeing and Hearing, Im2wav use VGGSound Dataset ($\sim$200,000 samples), and Diff-Foley uses AudioSet ($\sim$390,000 samples) in addition to VGGSound. ReelWave demonstrates effectiveness and potential to be further extended in larger datasets.

For temporal synchronization, ReelWave is ranked first on energy MAE, indicating its %our model can produce 
better structure prediction, as presented in Table~\ref{tab:temporal}. Although our model is ranked average in Onset detection, note that all the baselines are trained on the VGGSound dataset, which is a superset of AVSync. While Im2wav achieves excellent results on Onset Acc and AV-align, it is ranked last on the others. In summary, our work achieves balanced results across all the metrics, with our goal being to produce on- and off-screen sound. 

Indeed, we find ReelWave can achieve impressive results on musical instrument sound alignment. For such a kind of instrument sound, on-screen movement change can be subtle when the player does not move too much while performing. Since we incorporate pitch change as the condition, a property that can exhibit many changes in instrument videos, ReelWave can achieve more accurate alignment as the performer swiftly plays the next note. While Figures~\ref{fig:soundtrack} and~\ref{fig:scene_boundary} provide some relevant visualization, please see and listen to supplementary video and audio examples for comparison.

\begin{table}
\caption{Quantitative evaluation on semantic alignment and audio quality. Audio-Agent achieves on par performance versus state-of-the-art models in terms of Mean KL Divergence (MKL)~\cite{iashin2021taming}, CLIP~\cite{wu2022wav2clip} and FD~\cite{heusel2017gans} on AVSync15~\cite{zhang2024audio}. } 
\label{tab:semantic}
\vspace{-0.05in}
\label{tab.v2a_result}
\centering
\resizebox{0.75\linewidth}{!}{%
\begin{tabular}{lccc}
\toprule[1.5pt]
\multicolumn{1}{c}{Method} & \multicolumn{1}{c}{MKL  $\downarrow$} & \multicolumn{1}{c}{CLIP  $\uparrow$}  & \multicolumn{1}{c}{FD  $\downarrow$}   \\ 
\midrule[1.5pt]
Seeing and Hearing~\cite{xing2024seeing}             &        2.34       & 8.35   &59.23                                       \\
Im2wav~\cite{sheffer2023hear}       &  2.60      & 3.37  & 68.22  \\
Diff-Foley~\cite{luo2024diff}             &    2.13                  &  10.18 & 66.97    \\
Ours                &     1.85       & 9.58  & 60.39     
\\\bottomrule[1.5pt]
\end{tabular}
}
%\vspace{-0.15in}
\end{table}

\begin{table}
\caption{Quantitative evaluation on temporal synchronization. We report onset detection accuracy (Onset ACC) and average precision (Onset AP) for the generated audios on AVSync~\cite{zhang2024audio}, which provides onset timestamp labels for assessment, following previous studies~\cite{luo2024diff,xie2024sonicvisionlm}.}
\label{tab:temporal}
\vspace{-0.05in}
\label{tab.v2a_sync}
\centering
\resizebox{\linewidth}{!}{%
\begin{tabular}{lcccc}
\toprule[1.5pt]
\multicolumn{1}{c}{Method} & \multicolumn{1}{c}{Onset ACC $\uparrow$} & \multicolumn{1}{c}{Onset AP  $\uparrow$} & \multicolumn{1}{c}{MAE $\downarrow$} & \multicolumn{1}{c}{AV-Align $\uparrow$} \\ 
\midrule[1.5pt]
Seeing and Hearing~\cite{xing2024seeing}             & 29.80                    &    67.76   & 1.93   & 21.29                \\
Im2wav~\cite{sheffer2023hear}       &      35.13     & 62.12   & 2.78 & 24.72                                                     \\
Diff-Foley~\cite{luo2024diff}             &   25.50                     &    67.46   & 2.05    &  15.80      \\
Ours         &      25.99             &  65.62 & 1.92 & 21.16      \\
\bottomrule[1.5pt]
\end{tabular}
}
%\vspace{-0.20in}
\end{table}

\section{Conclusion and Discussion}
\label{sec:conclusion}

\noindent\textbf{Limitation and Future Work.~} ReelWave uses SynchFormer as the video feature encoder, which performs optimally when the input video does not exceed 5 seconds. While some previous methods use CLIP to extract per-frame features to get a longer feature sequence, it has been noted in~\cite{jeong2024read} that this method does not suit the temporal synchronization task. It is thus an important direction to conduct research in designing a better video feature extractor tailored for on-screen sound generation.
%In addition, ReelWave may be extended for lip synchronization and inferring dialogue from the talking-head video, given stronger video understanding model and the text-to-speech model. This may make a powerful impact in video-to-speech generation.

\vspace{2mm}
\noindent\textbf{Conclusion.~}
We present ReelWave, a multi-agent framework for multi-scene audio generation. Our key idea lies in decomposing the generation target to on-screen sound and off-screen sound, where the final audio is produced by overlapping soundtracks with appropriate volume and time occurrence to achieve rich content. Our work also proposes a methodology that uses time-varying interpretable control signals that are derived from the inherent audio. Quantitative results show that even without large-scale datasets, the Foley on-screen sound prediction model can achieve a promising result in terms of semantic alignment, audio quality, and temporal synchronization.

\newpage
\bibliographystyle{ACM-Reference-Format}
\bibliography{sample-base}

\begin{figure*}[t!]
    \centering
    \includegraphics[width=0.99\linewidth]{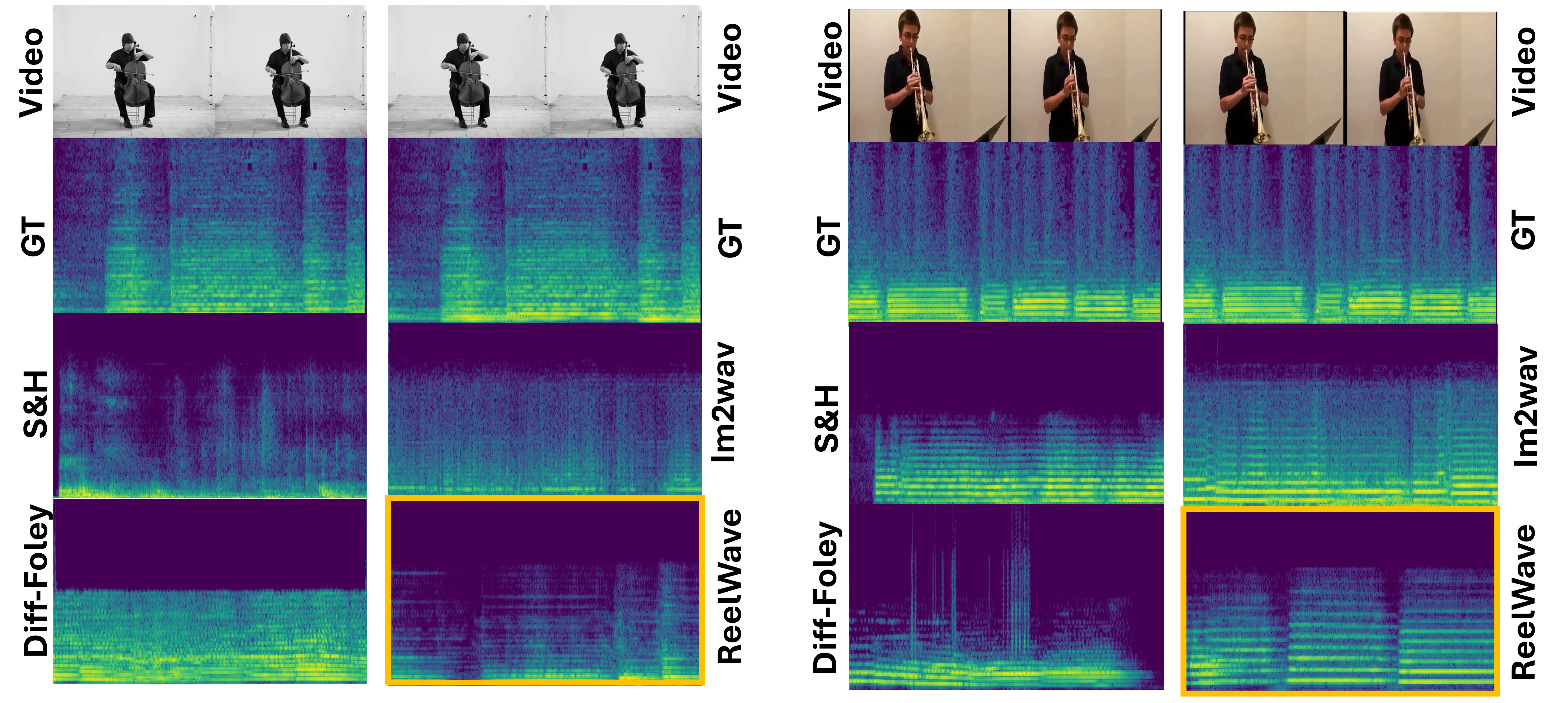}
    \caption{Qualitative comparison on AVSync dataset. ReelWave achieves clearer distinction between event occurrence especially on musical instrument sound alignment, compared to the baselines. To facilitate comparison, we arrange the spectrogram in two columns, given the same ground truth and video input. Our result is framed with a yellow border. Please see and listen to the supplemental videos with audio.}
    \label{fig:sync}
\end{figure*}

\vspace*{\fill}  % Push remaining content to the bottom

\begin{figure}[b!]
    \centering
    \includegraphics[width=0.99\linewidth]{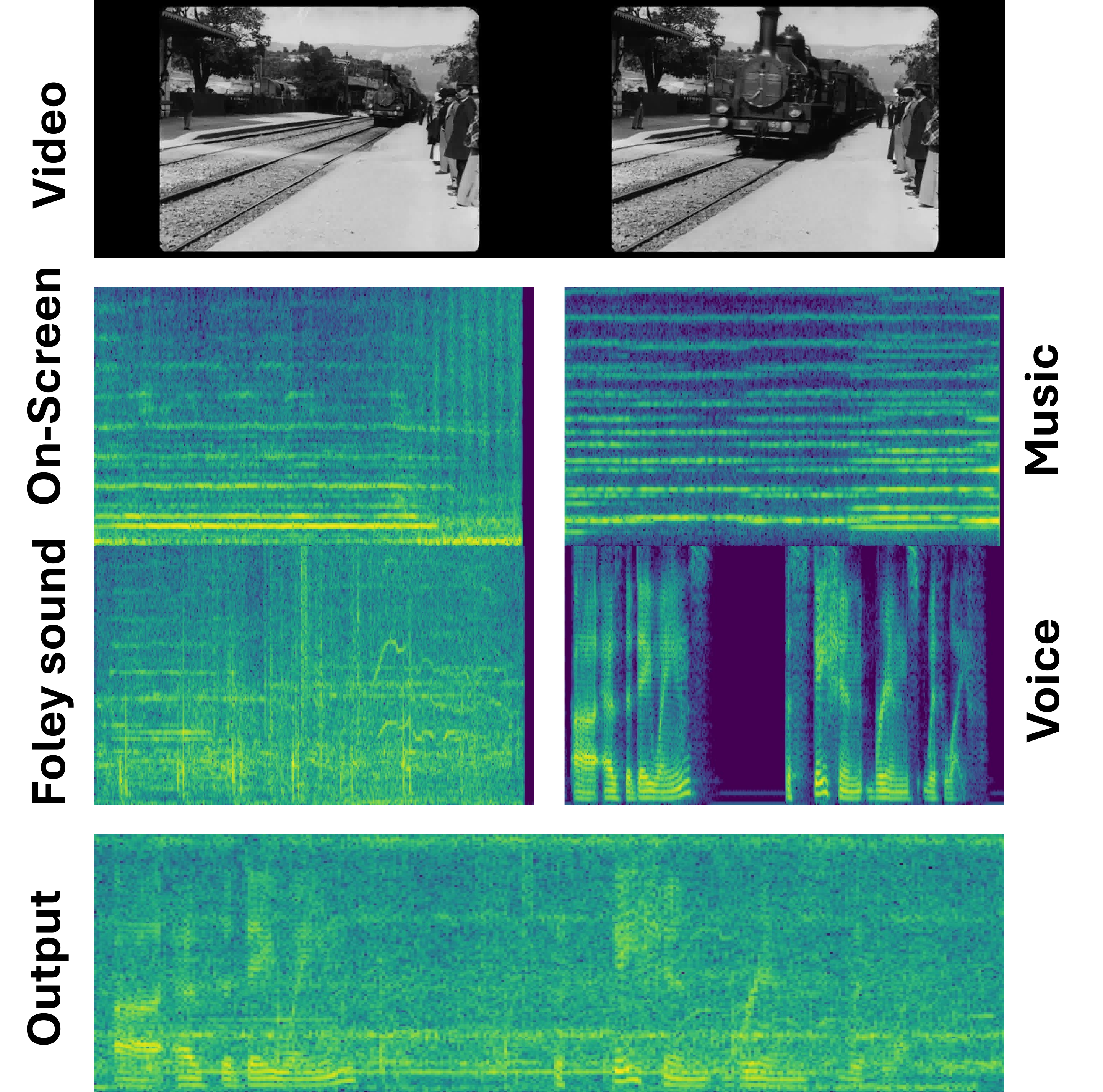}
    \caption{Visualizing the different \textbf{soundtracks} used to combine the final output. We reshape the output spectrogram to span the whole width of the figure; the length of the original output spectrogram is the same as other soundtracks.}
    \label{fig:soundtrack}
\end{figure}

\begin{figure}[b!]
    \centering
    \includegraphics[width=0.99\linewidth]{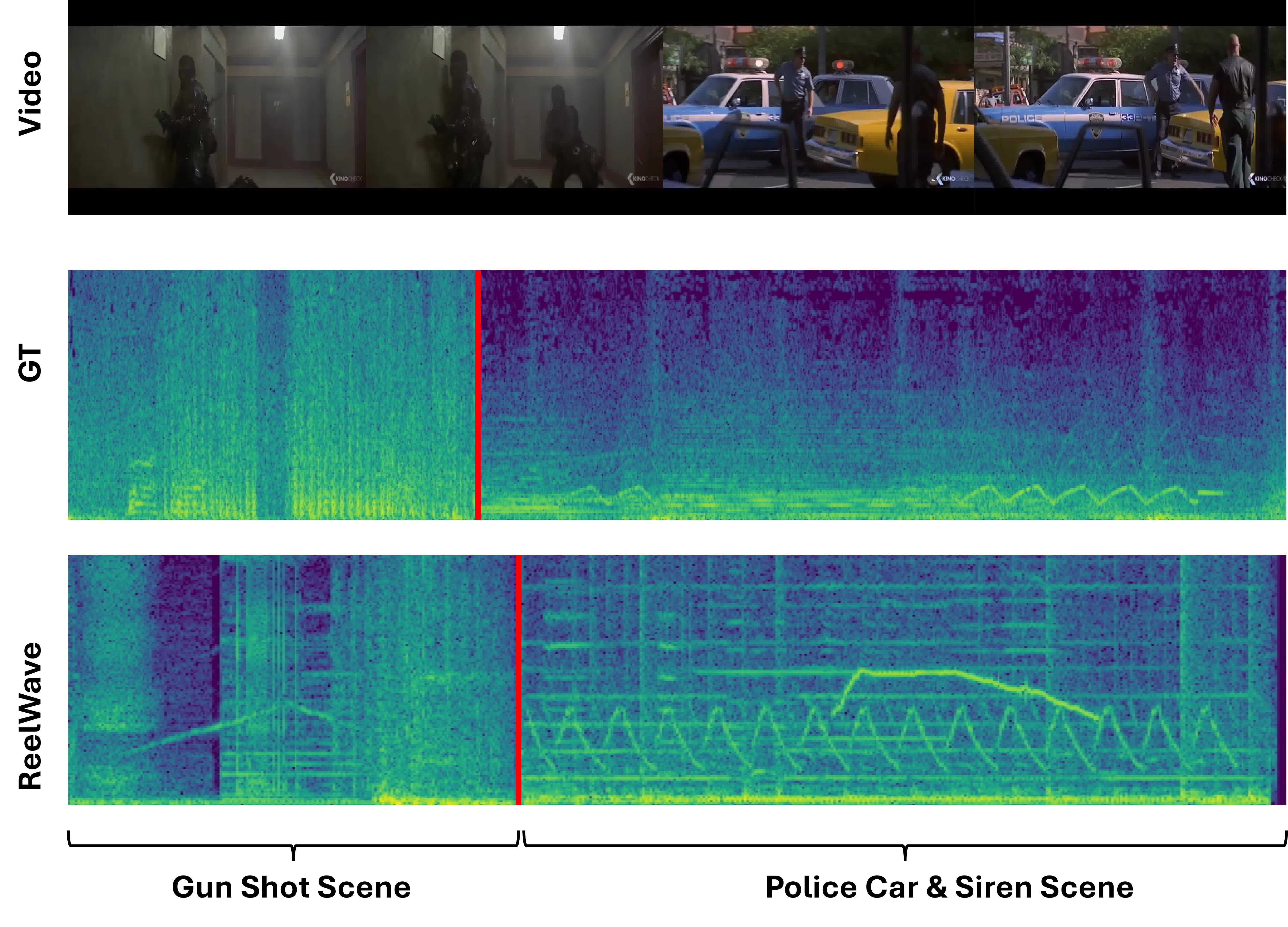}
    \caption{Visualizing the \textbf{scene boundary} given multi-scene video input. ReelWave can effectively recognize the distinction between scenes and generate soundtracks with suitable context, with almost perfect division of scene boundary.
        }
    \label{fig:scene_boundary}
\end{figure}

%%

%\vspace{2mm}

\clearpage
\appendix
\section{Appendix Section}
\label{sec:appendix_section}
\subsection{Script and generation plan example} 
We present the examples for the script, voice plan, mixer plan, Foley plan. See Figure ~\ref{fig:voice_plan},~\ref{fig:mixer_plan},~\ref{fig:foley_plan}, and~\ref{fig:music_plan}

\begin{figure}[b]
\begin{tcolorbox}[title=Voice Plan, fonttitle=\bfseries\small, coltitle=black, colbacktitle=gray!20, colback=white, colframe=gray!50, boxsep=1mm, left=1mm, right=1mm, top=1mm, bottom=1mm]
\begin{lstlisting}[
style=jsonstyle,
basicstyle=\footnotesize\ttfamily,
]
{
    "Scene": [
        {
            "ID": "1",
            "Layout": "foreground",
            "Sex": "male",
            "Dialogue": "We must save the town and end the violence. Hurry up.",
            "Start_time": 0,
            "End_time": 4
        }
    ]
}
\end{lstlisting}
\end{tcolorbox}

\caption{Voice plan example, which consists of ID, Layout, Sex, Dialogue, Start\_time, End\_time. The Layout can only be ``foreground" for clear hearing.}
\label{fig:voice_plan}
\end{figure}

\begin{figure}[b]
\begin{tcolorbox}[title=Mixer Plan, fonttitle=\bfseries\small, coltitle=black, colbacktitle=gray!20, colback=white, colframe=gray!50, boxsep=1mm, left=1mm, right=1mm, top=1mm, bottom=1mm]
\begin{lstlisting}[style=jsonstyle,
basicstyle=\footnotesize\ttfamily]
{
    "Scene": [
        {
            "Type": "On-screen",
            "ID": "1",
            "Start_time": 0,
            "End_time": 5,
            "Volume": -15
        },
        {
            "Type": "Foley",
            "ID": "1",
            "Start_time": 0,
            "End_time": 2,
            "Volume": -30
        },
        {
            "Type": "Voice",
            "ID": "1",
            "Start_time": 0,
            "End_time": 5,
            "Volume": -15
        },
        {
            "Type": "Music",
            "ID": "1",
            "Start_time": 0,
            "End_time": 5,
            "Volume": -35
        }
    ]
}
\end{lstlisting}
\end{tcolorbox}
\caption{Mixer plan example, which consists of Type, ID, Start\_time, End\_time, and Volume.}
\label{fig:mixer_plan}
\end{figure}

\begin{figure}[b]
\begin{tcolorbox}[title=Foley Plan, fonttitle=\bfseries\small, coltitle=black, colbacktitle=gray!20, colback=white, colframe=gray!50, boxsep=1mm, left=1mm, right=1mm, top=1mm, bottom=1mm]
\begin{lstlisting}[style=jsonstyle,
basicstyle=\footnotesize\ttfamily]
{
    "Scene": [
        {
            "ID": "1",
            "Layout": "background",
            "Description": "Distant river flowing",
            "Start_time": 0,
            "End_time": 3
        },
        {
            "ID": "2",
            "Layout": "foreground",
            "Description": "Projectiles whistling through the air",
            "Start_time": 0,
            "End_time": 2
        }
    ]
}
\end{lstlisting}
\end{tcolorbox}
\caption{Foley plan example, which consists of ID, Layout, Description, Start\_time, End\_time. The Layout can be foreground or background, which serves as a hint for the Mixer to assign appropriate volume.}
\label{fig:foley_plan}
\end{figure}

\begin{figure}[b]
\begin{tcolorbox}[title=Music Plan, fonttitle=\bfseries\small, coltitle=black, colbacktitle=gray!20, colback=white, colframe=gray!50, boxsep=1mm, left=1mm, right=1mm, top=1mm, bottom=1mm]
\begin{lstlisting}[style=jsonstyle,
basicstyle=\footnotesize\ttfamily]
{
    "Scene": [
        {
            "ID": "1",
            "Layout": "background",
            "Style": "Orchestral dramatic",
            "Description": "A grand orchestral arrangement with thunderous percussion, epic brass fanfares, and soaring strings, creating a cinematic atmosphere fit for the chaotic and urgent scene of a town under siege by fire and explosions.",
            "Start_time": 0,
            "End_time": 5
        }
    ]
}
\end{lstlisting}
\end{tcolorbox}
\caption{Music plan example, which consists of ID, Layout, Style, Description, Start\_time, End\_time. The Layout can only be ``background" to avoid being too loud and covering other sounds.}
\label{fig:music_plan}
\end{figure}

\end{document}